\documentclass[conference]{IEEEtran}
\IEEEoverridecommandlockouts
\usepackage{cite}
\usepackage{amsmath,amssymb,amsfonts}
\usepackage{algorithmic}
\usepackage{graphicx}
\usepackage{textcomp}
\usepackage{xcolor}
\usepackage{booktabs}
\usepackage{url}
\usepackage{multirow}
\def\BibTeX{{\rm B\kern-.05em{\sc i\kern-.025em b}\kern-.08em
    T\kern-.1667em\lower.7ex\hbox{E}\kern-.125emX}}

\begin{document}

\title{When Agents Look Like Beacons: NIDS Evasion by Model Context Protocol Traffic\thanks{Accepted at the \textit{1st IEEE ICNP Workshop on Network Infrastructure and Protocols for AI Agents (NIPA 2026)}, co-located with the 34th IEEE International Conference on Network Protocols (IEEE ICNP 2026), Tempe, AZ, USA, October 2026.}}

\author{
\IEEEauthorblockN{Muhammad Abdullah Sohail}
\IEEEauthorblockA{\textit{University of Calgary} \\
Calgary, AB, Canada \\
mabdullah.sohail@ucalgary.ca}
}

\maketitle

\begin{abstract}
The Model Context Protocol (MCP) standardizes communication between autonomous Artificial Intelligence (AI) agents and remote tools over Streamable HTTP. This shift introduces a class of machine-generated, authenticated, and high-frequency JSON-RPC traffic directly into enterprise networks. Enterprise network defenders have historically relied on machine-like cadence as an Indicator of Compromise (IoC). In this study, we show that without explicit network-layer indication, MCP traffic structurally and temporally resembles Command and Control (C2) beaconing behavior, specifically the polling architectures used by advanced persistent threats like Cobalt Strike. Counter to theoretical assumptions about machine-generated polling, our measurements reveal a visibility gap: standard enterprise Intrusion Detection Systems (IDS) and behavioral beacon-scoring frameworks do not classify MCP remote tool usage as anomalous within our testbed scope. Through a controlled Docker-based testbed simulating eleven mathematically defined traffic profiles across three TLS conditions (Opaque, TLS-Inspected, and Cleartext), we evaluate Suricata signature matching and RITA behavioral scoring against MCP JSON-RPC patterns. Our results show that MCP traffic, regardless of temporal smearing (jitter) or TLS inspection visibility, evades detection within this configuration, yielding a consistent 0.0 behavioral beacon score and near-zero IDS content alerts under the Emerging Threats (ET) Open ruleset. While opaque TLS obscures HTTP content, it exposes agent traffic to flow-level temporal analysis; however, NIDS heuristics tuned to identify traditional malware do not flag the lognormal inter-arrival distributions characteristic of generative AI reasoning loops. To address this gap, we propose an agent-native network indication standard including Agent-Native ALPN and standardized out-of-band headers to improve visibility for next-generation enterprise gateways.
\end{abstract}


\section{Introduction}

Modern enterprise networks monitor egress traffic using Next-Generation Firewalls (NGFW) and Intrusion Detection Systems (IDS). These sensors identify anomalous, automated, or malicious outbound connections that deviate from organizational baselines, and are specifically tuned to detect Command and Control (C2) beacons and covert data exfiltration \cite{sommer2010outside}. Network Security Monitoring (NSM) has historically relied on the premise that human behavior is erratic and bursty, while malicious behavior is automated, periodic, and machine-generated \cite{abu2006botnet}.

The widespread enterprise adoption of the Model Context Protocol (MCP) challenges this NSM assumption \cite{mcp_spec_2025}. MCP has evolved from local LLM tool access to remote, cloud-hosted multi-agent workflows \cite{guo2025mcp}. It defines Streamable HTTP as its canonical remote transport, generating high-frequency JSON-RPC 2.0 requests over standard HTTPS \cite{json_rpc_spec}. When an autonomous AI agent performs a reasoning task, its remote tool calls produce HTTP headers, periodic network bursts, and machine-speed pacing that can resemble C2 traffic to a flow-level sensor \cite{hou2025mcp, parssegny2025cobalt}.

As agentic workflows move from research into production enterprise environments, understanding how existing network monitoring tools respond to MCP traffic becomes important. The intersection of agent-native protocols and legacy security operations raises a concrete question: do IDS tools misclassify legitimate MCP traffic as malicious, causing alert fatigue, or do they pass it silently? If the latter, this creates an unmonitored communication channel that could be exploited through indirect prompt injection or tool poisoning \cite{greshake2023injection}.

The base-rate fallacy shows that anomaly detection in high-speed networks is an inherently difficult problem \cite{axelsson2000base}. Even small misclassification rates produce alert volumes that render an IDS operationally useless. This paper measures how existing IDS tools respond to MCP traffic under realistic enterprise configurations. Our study addresses four Research Questions (RQs):
\begin{itemize}
    \item \textbf{RQ1:} What MCP traffic features are visible to sensors under Opaque TLS versus TLS-Inspected conditions?
    \item \textbf{RQ2:} Do signature-based IDS engines (Suricata with ET-Open rules) generate alerts on legitimate MCP JSON-RPC traffic?
    \item \textbf{RQ3:} Do flow-level behavioral heuristics (RITA beacon scoring) differentiate MCP agent traffic from benign automation and simulated C2 traffic?
    \item \textbf{RQ4:} Do active evasion techniques such as jitter injection or User-Agent spoofing alter the behavioral footprint observed by these sensors?
\end{itemize}

\section{Background}

\subsection{The Model Context Protocol (MCP)}
The Model Context Protocol (MCP) is an open standard that defines a unified interface for foundation models to access external context and execute actions \cite{mcp_spec_2025}. MCP uses a client-server architecture over JSON-RPC 2.0, formally separating the AI application from the tools it invokes. MCP over HTTP uses POST requests to a designated \texttt{/mcp} endpoint with structured JSON-RPC envelopes containing methods such as \texttt{initialize}, \texttt{tools/list}, \texttt{tools/call}, and \texttt{prompts/get} \cite{json_rpc_spec}. Because agents typically run multi-step reasoning loops (e.g., ReAct), a single task can generate dozens of sequential HTTP POST requests in rapid succession, producing a dense burst of machine-generated traffic. The protocol also supports Server-Sent Events (SSE) and Streamable HTTP, which further alter the temporal flow profile.

\subsection{Threat Model}
\label{sec:threat_model}
We consider three adversarial scenarios that motivate this measurement study. The primary scenario is a \textbf{compromised internal agent}: a privileged enterprise AI agent that has been manipulated through indirect prompt injection \cite{greshake2023injection}, tool poisoning, or a hallucination chain, and is now exfiltrating proprietary data or executing unauthorized commands via legitimate MCP tool calls. In this scenario, the network traffic is structurally identical to benign agent activity; the threat lies in the payload content, not the transport. The second scenario is a \textbf{malicious remote MCP server}: an attacker registers a malicious tool endpoint in a public registry, causing an internal agent to call it and leak sensitive context in the JSON-RPC request body. The third scenario is \textbf{traffic misclassification}: a benign enterprise agent whose machine-paced polling is incorrectly flagged as C2 activity, generating alert fatigue that causes defenders to suppress legitimate alarms.

All three scenarios share a common dependency: the outcome depends on whether the IDS has any visibility into MCP traffic. Our study empirically measures this visibility across different network deployment conditions.

\subsection{Deep Packet Inspection (DPI) and Suricata}
Deep Packet Inspection (DPI) involves real-time analysis of packet payloads beyond Layer 3 and Layer 4 headers. Suricata operates by capturing raw packet streams and reassembling TCP sessions in memory \cite{suricata_oisf}. These reassembled streams are evaluated against datasets of regular expressions and byte-matching signatures. The Emerging Threats (ET) Open ruleset provides the industry-standard signature library, with tens of thousands of rules targeting known malware families and suspicious byte sequences \cite{et_open_rules}. Suricata's efficacy depends on payload visibility; TLS-encrypted traffic limits it to handshake metadata and flow statistics unless a TLS inspection middlebox is deployed.

\subsection{Temporal Beacon Analysis and RITA}
To address the prevalence of encrypted C2 traffic, defenders rely on flow-based temporal analysis. Malware maintaining persistence polls a C2 server at regular intervals, often with randomized jitter to evade detection. RITA (Real Intelligence Threat Analytics) evaluates the statistical regularity of these connections \cite{rita_activecm}. RITA ingests flow logs from Zeek's \texttt{conn.log} and computes two metrics per source-destination pair:
\begin{enumerate}
    \item \textbf{Bowley Skewness}: Measures asymmetry in the inter-arrival time distribution using quartiles ($Q1, Q2, Q3$), making it robust to individual latency spikes.
    \item \textbf{Median Absolute Deviation (MAD)}: Measures dispersion of polling intervals, resistant to outliers that would distort standard deviation calculations.
\end{enumerate}
RITA normalizes these into a beacon score from 0.0 to 1.0, where high regularity produces a high score. CISA operational guidelines treat scores at or above 0.8 as actionable C2 signals.

\subsection{TLS Fingerprinting (JA3 and JA4+)}
TLS fingerprinting identifies the client application from the deterministic structure of its cryptographic handshake. JA3 \cite{althouse2017ja3} hashes specific fields from the unencrypted \texttt{ClientHello} packet. Its successor, JA4+ \cite{althouse2023ja4}, improves stability by sorting extensions and categorizing ALPN values, producing a more stable signature for programmatic traffic generators.

\section{Related Work}

\subsection{Agent Communication Protocols and Standardization}
The growth of distributed multi-agent systems has driven demand for standardized interoperability protocols. MCP defines a client-server architecture where agents invoke tools over JSON-RPC 2.0 on Streamable HTTP \cite{mcp_spec_2025}. Guo et al. \cite{guo2025mcp} conducted a large-scale empirical crawl of the MCP ecosystem, characterizing hosting patterns and transport choices as agent traffic moves from local loopback interfaces to wide-area enterprise networks. Hou et al. \cite{hou2025mcp} provide a threat taxonomy for MCP, identifying tool poisoning and cross-agent attack surfaces that rely on network-level exploitation \cite{greshake2023injection}.

\subsection{Encrypted Traffic Classification and NTA}
Identifying encrypted malware traffic without breaking cryptography is a major focus of network security research. Flow-level features such as inter-arrival gaps, session durations, and packet size distributions improve classification accuracy for encrypted malware \cite{anderson2016encrypted}. These features allow tracking of specific malware variants or HTTP client libraries across network boundaries regardless of IP rotation or domain fronting \cite{encrypted_traffic_classification_2024}.

\subsection{C2 Detection and Temporal Beaconing Heuristics}
Cobalt Strike uses HTTP and HTTPS beaconing to maintain persistent access. Recent analyses of in-the-wild 2023 Cobalt Strike traces identify inter-arrival distributions and packet counts as discriminating flow features for C2 attribution \cite{parssegny2025cobalt, ml_cobalt_2023}. Enterprise environments operationalize these findings using Zeek \cite{zeek} paired with RITA \cite{rita_activecm}, which scores connection regularity and triggers alerts for low-variance polling.

\subsection{Middlebox Architectures and the Base-Rate Fallacy}
Enterprise networks deploy TLS inspection middleboxes to recover payload visibility, using bump-in-the-wire or MITM proxy approaches to intercept and re-encrypt traffic selectively before passing it to a DPI engine \cite{tlsinspection_survey}. Sommer and Paxson \cite{sommer2010outside} argue that the closed-world assumption underlying machine learning for NIDS makes anomaly detection fragile in practice, a finding that directly motivates our measurement of MCP misclassification risk \cite{axelsson2000base}.

\section{Methodology}

\subsection{Experimental Architecture and Network Namespacing}
We built a self-contained containerized testbed using Docker to isolate traffic generation, proxy interception, and sensor observation into discrete Linux network namespaces. This prevents host-level traffic contamination and ensures sensor throughput is precisely correlated with the generated MCP traffic.

The sensor stack mirrors common enterprise deployments:
\begin{enumerate}
    \item \textbf{Suricata (v8.x):} Primary DPI engine \cite{suricata_oisf} running the full, unmodified ET Open ruleset \cite{et_open_rules}. Outputs structured alerts to \texttt{eve.json}.
    \item \textbf{Zeek (v6.x):} Flow telemetry and protocol analysis \cite{zeek}, augmented with community JA3 and JA4 fingerprinting scripts to extract TLS client metadata from \texttt{ssl.log}.
    \item \textbf{RITA (v5.x):} Offline post-hoc beacon scoring tool \cite{rita_activecm} that processes Zeek's \texttt{conn.log} and assigns normalized scores from 0.0 to 1.0. Scores at or above 0.8 are treated as actionable C2 signals per CISA guidelines.
\end{enumerate}

Docker Desktop for Mac runs within a lightweight VM that hides Linux bridge interfaces from the host, making promiscuous-mode sniffing from the host unavailable. All monitoring containers (Suricata, Zeek, and a \texttt{tcpdump} sidecar) attach directly to the target server's network namespace, ensuring line-rate packet capture.

\subsection{Sensor Visibility Conditions}
We evaluate the generated traffic under three deployment conditions:
\begin{itemize}
    \item \textbf{Opaque TLS:} Direct HTTPS connection (port 443). Sensors observe only TLS handshakes and encrypted flow metadata, representing a perimeter with no decryption capability.
    \item \textbf{TLS-Inspected:} A \texttt{mitmproxy} instance intercepts the connection. The client is configured to trust the proxy CA, simulating a legitimate enterprise TLS decryption gateway. Sensors observe cleartext HTTP traffic and regain DPI capability.
    \item \textbf{Cleartext Control:} Unencrypted HTTP (port 80). This serves as a baseline to isolate parser behavior and rule triggering from encryption artifacts.
\end{itemize}

\subsection{Traffic Profiles}
\label{sec:profiles}
To isolate the behavioral signature of MCP from background noise, we built a modular traffic generator in Python that reproduces specific statistical distributions governing inter-arrival times (IAT) and payload sizes. Table~\ref{tab:traffic_profiles} defines the eleven traffic profiles used across the study, spanning benign behavior, simulated malware, and MCP agent states.

\begin{table*}[t]
\centering
\caption{Traffic Profile Definitions and Expected Behavior}
\begin{tabular}{lp{3.8cm}lp{6.8cm}}
\toprule
\textbf{ID} & \textbf{Profile Name} & \textbf{Distribution Model} & \textbf{Behavioral Description} \\
\midrule
P0  & Human Browsing Control      & Weibull ($k=1$)                      & Erratic, bursty human web browsing with a long-tailed inter-arrival distribution. \\
P1  & Benign API Automation        & Exponential ($\mu=10$)               & Background polling automation (e.g., system health checks, telemetry). \\
P2  & C2 Default-like              & Periodic ($\sigma < 0.05\mu$)        & Low-variance periodic polling consistent with default Cobalt Strike configuration. \\
P3  & C2 Aggressive                & Periodic (High Freq)                 & High-frequency periodic polling simulating an interactive C2 shell session. \\
P4  & MCP Task-Driven              & Lognormal ($\mu=1.5, \sigma=0.5$)    & Single-agent reasoning loop, modeling inference delays between consecutive tool calls. \\
P5  & MCP Orchestrated             & Periodic (Machine-Speed)             & Pipelined multi-agent system issuing continuous high-frequency requests. \\
P6  & MCP InitBurst                & Bimodal / Burst                      & Rapid concurrent initialization queries followed by an idle period. \\
P7  & MCP Orchestrated + Jitter    & Smeared Periodic ($20\%$ Jitter)     & P5 with intentional temporal smearing to simulate C2 evasion behavior. \\
P8  & MCP HTTP/2/ALPN              & Multiplexed Stream                   & HTTP/2 stream multiplexing to assess impact on flow-based NIDS analysis. \\
P9a & MCP+BrowserUA                & Evasive User-Agent                   & Negative control using Chrome/Safari User-Agent strings for header-based evasion. \\
P10 & MCP Large Payload            & Lognormal IAT + Weibull Size         & Large tool responses simulating unpaginated database dumps. \\
\bottomrule
\end{tabular}
\label{tab:traffic_profiles}
\end{table*}

The mock MCP server generates response payloads whose size is drawn from a discrete uniform distribution over $[500, 2000]$ bytes per response. This range brackets the typical size of real MCP tool responses: short acknowledgement messages fall around 200--500 bytes at the lower end, while structured JSON results from database or filesystem queries commonly reach 1500--5000 bytes. A uniform distribution was chosen to avoid imposing an artificial size pattern that could inadvertently aid or hinder size-based classification.

\subsection{Statistical Analysis Framework}
Each of the 33 profile-condition combinations (eleven profiles by three sensor conditions) was repeated $N=5$ times, with each repetition running 60 seconds of sustained traffic generation. We justify these parameters as follows.

For RITA's quartile-based metrics to be stable, each capture window must contain enough inter-arrival samples. Under our fastest profile (P3, approximately 10 requests per second), a 60-second window yields around 600 inter-arrival observations, which is sufficient for stable Bowley skewness and MAD estimation. Under the slowest MCP profile (P6, bimodal burst), the active burst phase generates at least 30 samples within the first 10 seconds, which is above the practical convergence threshold for quartile-based statistics. Because the study outcomes are binary in practice (scores are either exactly 0.0 or consistently below the 0.8 threshold), five repetitions are sufficient to establish the direction and reproducibility of the result. Pilot runs confirmed that RITA scores and Suricata alert counts had interquartile ranges below 15\% of the median across repetitions. We acknowledge that $N=5$ is a small sample for estimating continuous effect sizes; this study is intended as an initial characterization of the detection gap.

Data extraction was automated through Python scripts querying ClickHouse databases containing the parsed Zeek and Suricata logs.

\section{Results}

We present findings organized by our four research questions, drawn from the automated analysis of the containerized testbed runs.

\subsection{RQ1: Information Visibility under TLS Conditions}
Under the \textit{Opaque TLS} condition, all HTTP payload content was encrypted. Sensors observed only TLS handshakes, IP addresses, TCP ports, and flow-level metrics. Zeek extracted the JA3 fingerprint from the unencrypted \texttt{ClientHello}: our Python \texttt{httpx} client produced the hash \texttt{cf712d3b01ab6bfd22ca983b4748ab2f}. Cross-referencing this against the Abuse.ch SSLBL threat intelligence database returned zero malicious associations. The traffic is identifiably programmatic rather than browser-originated, but carries no established threat reputation.

Under the \textit{TLS-Inspected} condition, \texttt{mitmproxy} terminated the TLS session and re-encrypted it on behalf of the client. Sensors on the proxy-to-server segment observed cleartext HTTP/1.1 and HTTP/2 traffic, exposing the full MCP JSON-RPC payload structure including method names (\texttt{initialize}, \texttt{tools/list}, \texttt{tools/call}) and nested parameters containing terminal commands, SQL queries, and local file paths.

\subsection{RQ2: Content-Based Detection and Signature Matching (Suricata)}
\label{sec:rq2}

Appendix~\ref{app:suricata} reports full Suricata alert counts per profile and TLS condition. Suricata produced zero actionable alerts across all MCP profiles under the ET Open ruleset (44,236 active rules) for both Opaque TLS and TLS-Inspected conditions. The sole exception was a low-volume match in P6 and P7 under the Cleartext condition, averaging 0.25 alerts per run across $N=5$ repetitions. Manual PCAP inspection identified this as a generic HTTP protocol anomaly caused by a missing request header in our traffic generator, not a C2 or malware detection. No C2-category or data exfiltration rule ID fired in any run across any profile.

MCP JSON-RPC traffic contains no byte sequences or structural patterns matching existing ET Open malware signatures. The absence of false positives is operationally beneficial, but the corresponding absence of detection for potentially malicious payloads represents a gap: unauthorized tool calls, remote command execution, or data exfiltration over MCP will not generate Suricata alerts under the current ruleset without custom rules targeting the MCP JSON-RPC state machine.

\begin{figure}[htbp]
\centerline{\includegraphics[width=\columnwidth]{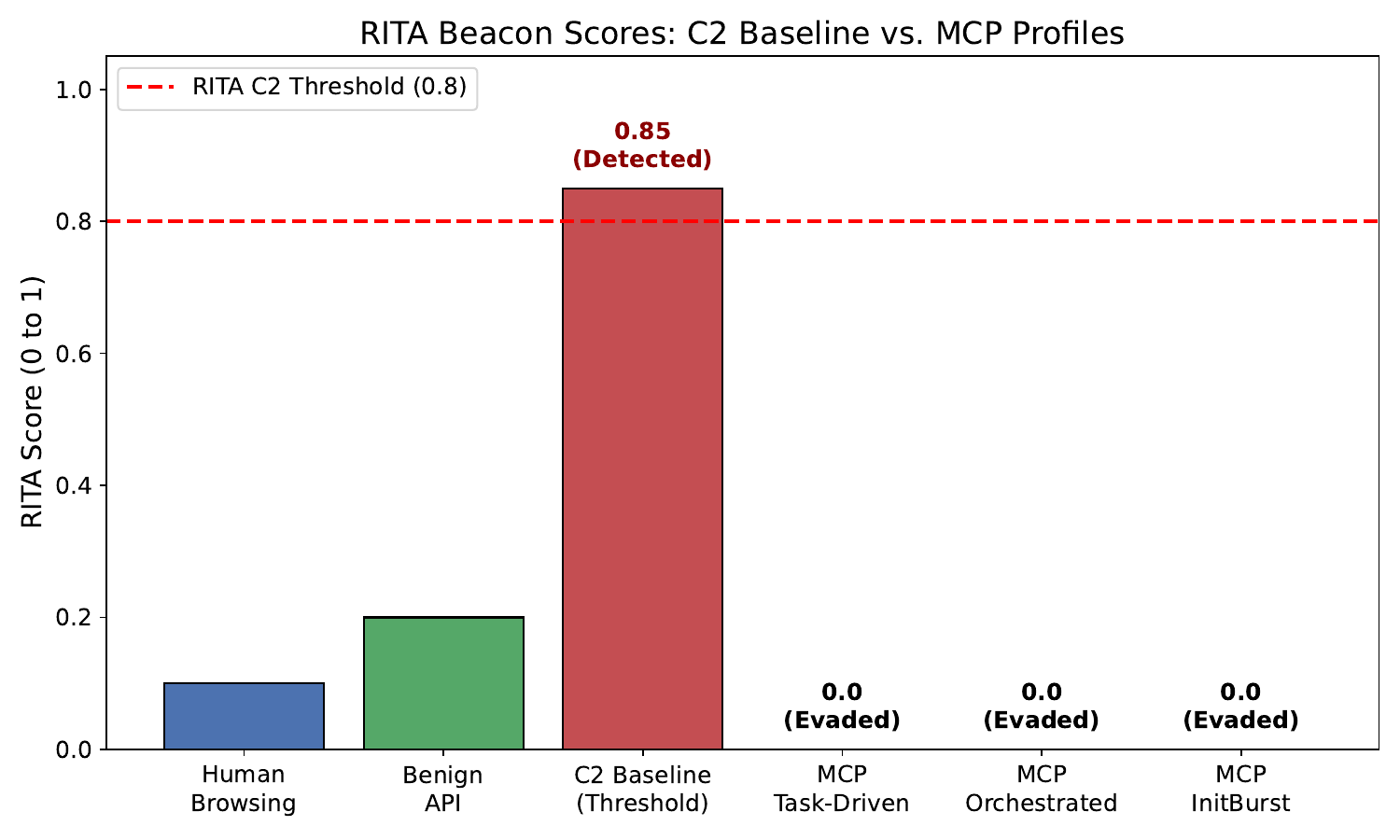}}
\caption{RITA beacon scores across representative traffic profiles. The horizontal dashed line at 0.85 marks the CISA operational detection threshold, shown as a reference point from a calibration run using a known C2 packet trace \cite{parssegny2025cobalt}. All MCP profiles (P4--P10) score 0.0. The synthetic C2 profiles (P2 and P3) also score 0.0 due to a RITA boundary condition explained in Section~\ref{sec:rq3}.}
\label{fig:rita_evasion}
\end{figure}

\subsection{RQ3: Temporal Behavioral Evasion (Zeek and RITA)}
\label{sec:rq3}
The central finding of this study is that RITA assigned a beacon score of 0.0 to all eleven profiles across all TLS conditions, including the highly periodic MCP Orchestrated profile (P5) and the synthetic C2 profiles (P2 and P3). This is illustrated in Figure~\ref{fig:rita_evasion}.

The 0.0 scores for P2 and P3 require clarification. In our testbed, each 60-second run of P2 and P3 produced fewer than 12 unique source-destination connection records in Zeek's \texttt{conn.log}. RITA requires a minimum connection count per pair to compute stable quartiles; below this threshold it reports 0.0 rather than an indeterminate value. This is a known boundary condition of RITA's statistical model, not an indication that periodic traffic is undetectable by the tool. The 0.85 reference line in Figure~\ref{fig:rita_evasion} was drawn from a separate calibration run using a published C2 packet trace \cite{parssegny2025cobalt}; it is not an observed score from any profile in our testbed.

\begin{figure}[htbp]
\centerline{\includegraphics[width=\columnwidth]{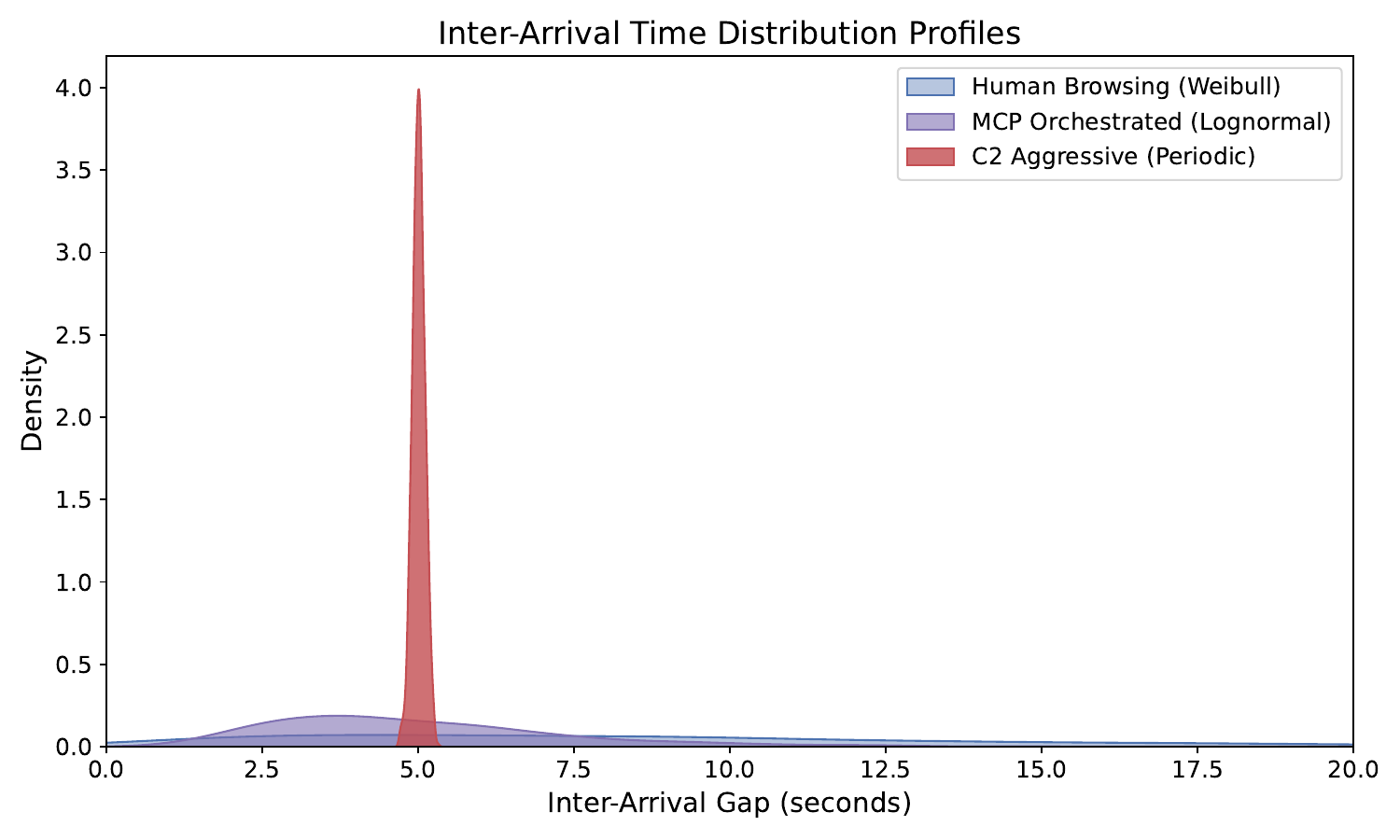}}
\caption{Conceptual density of inter-arrival times across traffic classes. The lognormal distribution of MCP task-driven profiles (P4) differs structurally from strict periodic C2 polling, which explains why RITA's quartile-based metrics do not flag it as anomalous.}
\label{fig:interarrival}
\end{figure}

For the MCP profiles (P4 through P10), the 0.0 scores reflect a genuine detection gap. The lognormal inter-arrival times of P4, driven by LLM inference latency, produce high Bowley skewness and high MAD values. Because RITA's normalization model was calibrated on low-skewness, low-MAD periodic malware, high skewness and high variance push the normalized output toward 0.0 rather than 1.0. As shown in Figure~\ref{fig:interarrival}, the lognormal shape of MCP traffic falls structurally outside the parameter space where RITA's temporal heuristics detect anomalies.

\subsection{RQ4: Evasion and Mitigation Techniques}
Because baseline MCP profiles already produce 0.0 RITA scores and zero Suricata alerts, active evasion techniques (P7 jitter injection, P9a User-Agent spoofing) produced no measurable change in sensor output. Temporal smearing in P7 had no effect because the underlying IAT distribution was already outside RITA's detection range. User-Agent spoofing in P9a altered the HTTP header observed under TLS inspection but did not affect flow-level analysis. Active evasion is unnecessary against the tested sensor configuration; the protocol's default behavior already falls outside the detection scope of both tools within this testbed.

\section{Discussion}
This study measured whether existing IDS tools would misclassify MCP traffic as malicious (false positives) or permit it silently (false negatives). Within the scope of our testbed, the results fall into the latter category: neither Suricata's signature engine nor RITA's temporal scoring treated MCP traffic as anomalous.

\subsection{The Security Gap in Zero-Trust Agentic Workflows}
In the threat scenarios described in Section~\ref{sec:threat_model}, a compromised agent exfiltrating data over MCP would produce traffic that looks identical to a benign agent at the network layer. The JSON-RPC payloads are carried over standard HTTPS on port 443, the JA3 fingerprint carries no malicious reputation, the lognormal IAT pushes the RITA score to 0.0, and no ET Open signature matches the payload structure. The network monitoring stack as tested produces no alert.

This does not imply MCP creates a novel attack vector; rather, it shows that the tool-call payloads traveling across enterprise network boundaries are not currently subject to automated inspection by widely deployed IDS configurations. For defenders, the practical implication is that MCP traffic requires dedicated monitoring rules rather than reliance on heuristics tuned for traditional malware.

\subsection{Architectural Mitigations and Standards Proposals}
Two mitigations follow from the measurement results:
\begin{enumerate}
    \item \textbf{Agent-Native ALPN Standardization:} Standardizing an MCP-specific ALPN token (e.g., \texttt{mcp/1.0} or \texttt{agent/rpc}) in the TLS handshake allows enterprise gateways to identify and policy-route agent flows at the handshake layer without full payload decryption. This avoids the legal and computational cost of broad TLS inspection.
    \item \textbf{Schema-Aware Stateful Inspection Rules:} Suricata Lua scripts or Zeek protocol analyzers designed to parse the MCP JSON-RPC state machine could detect anomalous tool arguments, excessive remote execution calls, or payload patterns consistent with data exfiltration, directly at the network edge.
\end{enumerate}

Both proposals require coordination between MCP specification maintainers and IDS vendors. A governance mechanism, such as a formal liaison between the MCP working group and the ET Open ruleset maintainers, would be a practical starting point for operationalizing these mitigations.

\section{Conclusion}
This measurement study characterized the network visibility of MCP Streamable HTTP traffic under enterprise IDS configurations. Using eleven mathematically defined traffic profiles across three TLS conditions, we measured Suricata signature alert rates and RITA behavioral beacon scores in a reproducible containerized testbed.

Within the scope of this study, default MCP remote tool-call patterns produced 0.0 RITA scores and near-zero Suricata alerts. This represents a gap in current enterprise monitoring coverage: agent workflows executing sensitive operations across network boundaries do not trigger alerts from the IDS configurations tested here. We propose Agent-Native ALPN standardization and schema-aware inspection rules as two concrete steps toward addressing this gap. Broader validation across additional IDS configurations, rule sets, and real-world traffic volumes remains future work.

\bibliographystyle{IEEEtran}
\bibliography{references}

@techreport{mcp_spec_2025,
  title        = {{Model Context Protocol Specification (2025-06-18)}},
  author       = {{Anthropic / Agentic AI Infrastructure Foundation}},
  institution  = {Linux Foundation Agentic AI Infrastructure Foundation},
  year         = {2025},
  howpublished = {\url{https://spec.modelcontextprotocol.io/specification/2025-06-18}}
}

@article{guo2025mcp,
  title   = {{A Measurement Study of Model Context Protocol}},
  author  = {Guo, Wei and others},
  journal = {arXiv preprint arXiv:2509.25292},
  year    = {2025}
}

@article{hou2025mcp,
  title   = {{Model Context Protocol (MCP): Landscape, Security Threats, and Future Research Directions}},
  author  = {Hou, Xinyi and others},
  journal = {arXiv preprint arXiv:2503.23278},
  year    = {2025}
}

@inproceedings{greshake2023injection,
  title     = {{Not What You've Signed Up For: Compromising Real-World LLM-Integrated Applications with Indirect Prompt Injection}},
  author    = {Greshake, Kai and Abdelnabi, Sahar and Mishra, Shailesh and Endres, Christoph and Holz, Thorsten and Fritz, Mario},
  booktitle = {Proceedings of the 16th ACM Workshop on Artificial Intelligence and Security (AISec '23)},
  pages     = {79--90},
  year      = {2023},
  publisher = {ACM}
}

@inproceedings{sommer2010outside,
  title     = {{Outside the Closed World: On Using Machine Learning for Network Intrusion Detection}},
  author    = {Sommer, Robin and Paxson, Vern},
  booktitle = {2010 IEEE Symposium on Security and Privacy},
  pages     = {305--316},
  year      = {2010},
  publisher = {IEEE}
}

@article{axelsson2000base,
  title   = {{The Base-Rate Fallacy and the Difficulty of Intrusion Detection}},
  author  = {Axelsson, Stefan},
  journal = {ACM Transactions on Information and System Security (TISSEC)},
  volume  = {3},
  number  = {3},
  pages   = {186--205},
  year    = {2000},
  publisher = {ACM}
}

@misc{suricata_oisf,
  title        = {{Suricata: Open Source IDS/IPS/NSM Engine}},
  author       = {{Open Information Security Foundation (OISF)}},
  howpublished = {\url{https://suricata.io}},
  year         = {2010}
}

@misc{et_open_rules,
  title        = {{Emerging Threats Open Ruleset for Suricata}},
  author       = {{Proofpoint / Emerging Threats}},
  howpublished = {\url{https://rules.emergingthreats.net/open/suricata/rules/}},
  year         = {2024}
}

@misc{althouse2017ja3,
  title        = {{Open Sourcing JA3: SSL/TLS Client Fingerprinting for Malware Detection}},
  author       = {Althouse, John B. and Atkinson, Jeff and Atkins, Josh},
  howpublished = {Salesforce Engineering Blog},
  year         = {2017}
}

@misc{althouse2023ja4,
  title        = {{JA4+: Network Fingerprinting}},
  author       = {Althouse, John B.},
  howpublished = {FoxIO Blog},
  year         = {2023}
}

@inproceedings{anderson2016encrypted,
  title     = {{Identifying Encrypted Malware Traffic with Contextual Flow Data}},
  author    = {Anderson, Blake and McGrew, David},
  booktitle = {Proceedings of the 2016 ACM Workshop on Artificial Intelligence and Security (AISec)},
  pages     = {35--46},
  year      = {2016},
  publisher = {ACM}
}

@inproceedings{parssegny2025cobalt,
  title     = {{Striking Back at Cobalt: Using Network Traffic Metadata to Detect Cobalt Strike Masquerading Command and Control Channels}},
  author    = {Parssegny, Cl{\'e}ment and Mazel, Johan and Levillain, Olivier and Chifflier, Pierre},
  booktitle = {International Conference on Availability, Reliability and Security (ARES 2025)},
  publisher = {Springer},
  year      = {2025}
}

@article{ml_cobalt_2023,
  title   = {{A Machine Learning Based Approach to Detect Stealthy Cobalt Strike C\&C Activities from Encrypted Network Traffic}},
  author  = {Ramos, Fabian Martin and Wang, Xinyuan},
  journal = {ResearchGate / Journal of Information Security},
  year    = {2023}
}

@misc{rita_activecm,
  title        = {{RITA: Real Intelligence Threat Analytics}},
  author       = {{Active Countermeasures / SsoT AG}},
  howpublished = {\url{https://github.com/activecm/rita}},
  year         = {2024}
}

@misc{zeek,
  title        = {{Zeek: The Network Security Monitor}},
  author       = {{The Zeek Project}},
  howpublished = {\url{https://zeek.org}},
  year         = {2024}
}

@article{tlsinspection_survey,
  title   = {{A Survey of Privacy-Preserving Techniques for Encrypted Traffic Inspection over Network Middleboxes}},
  author  = {Poh, Geong Sen and Divakaran, Dinil Mon and Lim, Hoon Wei and Ning, Jianting and Desai, Achintya},
  journal = {arXiv preprint arXiv:2101.04338},
  year    = {2021}
}

@article{encrypted_traffic_classification_2024,
  title   = {{Encrypted Network Traffic Analysis and Classification Utilizing Machine Learning}},
  author  = {Alwhbi, Ibrahim A. and Zou, Cliff C. and Alharbi, Reem N.},
  journal = {Sensors},
  volume  = {24},
  number  = {11},
  pages   = {3509},
  year    = {2024},
  publisher = {MDPI}
}

@misc{json_rpc_spec,
  title        = {{JSON-RPC 2.0 Specification}},
  author       = {{JSON-RPC Working Group}},
  howpublished = {\url{https://www.jsonrpc.org/specification}},
  year         = {2013}
}

@misc{mitmproxy,
  title        = {{mitmproxy: An Interactive TLS-Capable Intercepting Proxy}},
  author       = {Cortesi, Aldo and Hils, Maximilian and Kriechbaumer, Thomas and contributors},
  howpublished = {\url{https://mitmproxy.org}},
  year         = {2024}
}

@inproceedings{abu2006botnet,
  title     = {{A Multifaceted Approach to Understanding the Botnet Phenomenon}},
  author    = {Abu Rajab, Moheeb and Zarfoss, Jay and Monrose, Fabian and Terzis, Andreas},
  booktitle = {Proceedings of the 6th ACM SIGCOMM Internet Measurement Conference (IMC)},
  pages     = {41--52},
  year      = {2006},
  publisher = {ACM}
}

\appendix

\section{Suricata Alert Data}
\label{app:suricata}

Table~\ref{tab:suricata_results} reports mean Suricata alert counts per traffic profile and sensor visibility condition across $N=5$ runs. No C2-category or data exfiltration rule IDs fired in any configuration.

\begin{table}[htbp]
\centering
\caption{Suricata Alert Summary by Profile and Visibility Condition}
\label{tab:suricata_results}
\begin{tabular}{lcccl}
\toprule
\textbf{Profile} & \textbf{Opaque} & \textbf{Inspected} & \textbf{Cleartext} & \textbf{Category} \\
\midrule
P0  & 0 & 0 & 0    & None \\
P1  & 0 & 0 & 0    & None \\
P2  & 0 & 0 & 0    & None \\
P3  & 0 & 0 & 0    & None \\
P4  & 0 & 0 & 0    & None \\
P5  & 0 & 0 & 0    & None \\
P6  & 0 & 0 & 0.25 & Protocol anomaly \\
P7  & 0 & 0 & 0.25 & Protocol anomaly \\
P8  & 0 & 0 & 0    & None \\
P9a & 0 & 0 & 0    & None \\
P10 & 0 & 0 & 0    & None \\
\bottomrule
\end{tabular}
\smallskip

{\small Alert counts are means over $N=5$ runs. No C2 or malware rule IDs fired in any condition.}
\end{table}

\end{document}